# Physics-Aligned Deep Learning Enables SERS Resolving and Sequencing of Dynamic Single-Molecule DNA Oligomers in Plasmonic Nanocavity

Kuo Zhan[1,3], Peilin Xin[1,3], Yingqi Zhao[1,3], Han Gu[4], Enock Adjei Agyekum[5], Zhou Chen[4], Shuai Li[5], Jian Ye[4], Lu Cheng[6], Jian-an Huang[1,2,3]*

[1] Research Unit of Health Sciences and Technology, Faculty of Medicine, University of Oulu,

Aapistie 5 A, 90220 Oulu, Finland.

[2] Research Unit of Disease Networks, Faculty of Biochemistry and Molecular Medicine,

University of Oulu, Aapistie 5 A, 90220 Oulu, Finland.

[3] Biocenter Oulu, University of Oulu, Aapistie 5 A, 90220 Oulu, Finland.

[4] State Key Laboratory of Systems Medicine for Cancer, School of Biomedical Engineering, Shanghai Jiao Tong University, Shanghai, People's Republic of China

[5] The Biomimetics and Intelligent Systems (BISG) research unit, Faculty of Information Technology and Electronic Engineering, University of Oulu, Oulu, Finland

[6] Institute of Biomedicine, School of Medicine, Faculty of Health Sciences, University of Eastern Finland, Kuopio, Finland

*Email: jianan.huang@oulu.fi

## Abstract

Single-molecule surface-enhanced Raman spectroscopy (SM-SERS) captures dynamic molecular behavior with ultrahigh sensitivity, but its biopolymer analysis is hindered by strong spectral heterogeneity, transient hotspot sampling, and background interference. Here, we develop a physics-aligned deep learning framework integrating contrastive attention-based multiple-instance learning (CAMIL), a tri-channel multi-kernel CNN classifier, and trajectory-level transition-guided sequence reconstruction to decode single-molecule DNA oligomer dynamics in a plasmonic nanocavity. In this work, CAMIL mines informative spectra enriched with chain-embedded nucleotide and dinucleotide signatures from contrastive positive and negative DNA trajectory bags, generating a single-molecule DNA-segment spectral-states library. This library trains a 12-class CNN to assign query time-resolved DNA SM-SERS frames to high-confidence DNA-segment states, enabling trajectory-level analysis of state composition, dwell length, entropy, switching frequency, and transition matrix. The transition matrix is further converted into state-transition-edge evidence for candidate sequence scoring, enabling asymmetric DNA sequences inference from confined stochastic sampling dynamics. This framework transforms SM-SERS heterogeneity into quantitative analytical information, advancing dynamic single-molecule decoding.

## INTRODUCTION

Single-molecule surface-enhanced Raman spectroscopy (SM-SERS) offers an ultrasensitive approach for probing molecular behavior at plasmonic interfaces(*1–3*). By concentrating electromagnetic fields into nanoscale hotspots, SM-SERS enables the observation of single-molecular adsorption, conformational dynamics, and nanoscale interfacial interactions that are otherwise obscured in ensemble-averaged

measurements(*4–6*). Recently, we developed a plasmonic nanocavity based particle-in-well (PIW) platform for time-resolved SM-SERS measurements of DNA oligomers adsorbed on gold nanoparticles(*7*). In this architecture, a DNA-adsorbed gold nanoparticle is trapped within a gold-nano-well in air, forming a fixed plasmonic nanocavity hotspot at the particle–wall junction with strong electromagnetic fields. This fixed hotspot enables time-resolved monitoring of the random diffusion of individual DNA molecules on the nanoparticle surface with a spatial resolution corresponding to 1~2 DNA bases (Figure 1A). Related localized SM-SERS sensing platforms have also advanced biopolymer analysis (*8*, *9*) and emerging technologies related to mRNA, DNA, and protein sequencing (*10–12*).

However, the local-field confinement that enables single-molecule sensitivity also imposes an analytical limitation. As the hotspot samples only 1~2 DNA bases at a time, different frames may probe different local segments of the same DNA oligomer and generate distinct DNA-segment spectral states. Even for the same nominal segment, spectra can still vary from frame to frame. This spectral variability can be understood from the electromagnetic framework of SM-SERS, in which the intensity of the i-th vibrational mode from the hotspot-sampled molecule can be expressed as (*13–15*):

$$I_{\mathrm{SM-SERS},i} \propto \left|\mathbf{E}_{\mathrm{loc}}\left(\mathbf{r}_{m}, \omega_{R,i}\right) \cdot \mathbf{G}(\Omega)\, \boldsymbol{\alpha}_i'\, \mathbf{G}^{-1}(\Omega) \cdot \mathbf{E}_{\mathrm{loc}}(\mathbf{r}_{m}, \omega_0)\right|^2, \quad (1)$$

where $\mathbf{E}_{\mathrm{loc}}(\mathbf{r}_m, \omega_0)$ is the local electric field at the molecular position $\mathbf{r}_m$ at the excitation frequency $\omega_0$, $\mathbf{E}_{\mathrm{loc}}(\mathbf{r}_m, \omega_{R,i})$ is the local field at the Raman-scattered frequency $\omega_{R,i}$, $\boldsymbol{\alpha}_i'$ is the Raman tensor of the $i$-th vibrational mode, and $\mathbf{G}(\Omega)\boldsymbol{\alpha}_i'\mathbf{G}^{-1}(\Omega)$ represents the Raman tensor transformed from the molecular frame into the laboratory frame according to the molecular orientation Ω.

According to this expression, the observed SM-SERS signal is governed by the coupled effects of the excitation-side local field(*15*, *16*), the emission-side plasmonic enhancement(*17*, *18*), the molecular position within the hotspot(*19*), and the orientation-dependent Raman tensor projection(*20*, *21*). This intrinsic heterogeneity also limits the direct use of simple molecular standards as one-to-one spectral references for single-molecule DNA analysis. Isolated nucleobases, nucleosides, nucleotides, or dinucleotide standards do not fully reproduce the chemical and structural environment of bases embedded within an intact single DNA chain. In a single DNA oligomer, each base is covalently connected through the sugar-phosphate backbone and is influenced by base stacking, local chain conformation, charge distribution, and transient surface-contact geometry within the plasmonic hotspot. As a result, the SM-SERS fingerprints of chain-embedded nucleotides and dinucleotides can differ substantially from those of isolated A, T, C, G units.

Beyond confined stochastic local readout and frame-to-frame spectral variability, citrate ligands used to stabilize colloidal nanoparticles can coexist with DNA within the hotspot and contribute overlapping spectral features. Recent studies also suggest that ultrasmall plasmonic nanogaps can support molecular infiltration and exchange(*22*), indicating that spectral fluctuations may reflect not only molecular identity and conformation, but also transient hotspot occupancy and dynamic adsorbate redistribution. As a result, time-resolved SM-SERS measurements of single-molecule DNA oligomers generate complex spectral trajectories in which only a subset of frames contains clear, molecularly meaningful vibrational fingerprints(*4*, *7*, *23*). Taking SM-SERS of ATCGACTG as an example, the time-resolved SERS map shows strong frame-to-frame fluctuations rather than a static molecular fingerprint (Fig. 1B). Representative spectra from the same trajectory reveal AC-, TC&Cit-, and AT-dominated states (Fig. 1C–E), indicating that each frame reports a localized molecular readout arising from a local dinucleotide unit, nearby backbone contributions, or citrate within the hotspot. Thus, the key challenge in dynamic SM-SERS is not signal detection alone, but extracting molecularly meaningful information from heterogeneous, sparse, and transient spectral observations.

Here, we introduce a physics-aligned contrastive attention-based multiple-instance learning method (CAMIL) that connects spectral mining with the confined sampling physics of plasmonic nanocavity to identify dynamic hotspot-sampled chain-embedded DNA segment of nucleotide and dinucleotide signatures. Multiple-instance learning (MIL) is a weakly supervised strategy for learning from ambiguously labeled data (*24–26*), in which labels are assigned to groups of observations, or bags, rather than to individual instances. This approach is well suited to SM-SERS DNA trajectory datasets acquired in plasmonic nanocavities, because each trajectory set can be treated as a bag of stochastic local readouts. Individual spectra within each bag serve as instances and may contain sparse and unevenly distributed nucleotide-, dinucleotide-, or citrate-dominated vibrational information. In this study, CAMIL is used as a contrastive spectral-mining strategy. SM-SERS trajectory sets from the DNA oligomers containing the target chain-embedded nucleotides or dinucleotides are defined as positive bags, whereas trajectory sets from structurally related oligomers that share common chemical components but lack these targets are used as negative bags (Table 1). This contrastive design could suppress shared spectral contributions while selectively enriching non-overlapping, nucleotide- or dinucleotide-specific vibrational features. In this way, physically meaningful chain-embedded DNA segments signatures can be mined directly from SM-SERS DNA trajectories without requiring manual frame-level annotation.

Beyond identifying representative nucleotide- and dinucleotide-dominated spectra embedded within the DNA oligomer, it is also essential to decode how DNA oligomers stochastically diffuse and switch among DNA-segment spectral states at hotspot over time. In a plasmonic nanocavity, each DNA SM-SERS trajectory arises from a series of local sampling events, during which different DNA segments enter the enhanced field, adopt distinct adsorption geometries and orientations, remain in the hotspot for finite durations, and compete with coexisting surface species such as citrate. Tracking these DNA-segment state transitions converts SM-SERS heterogeneity into dynamic molecular information by revealing how local DNA segments are accessed, sampled and connected through sequence-dependent transitions. This dynamic view is conceptually aligned with single-molecule sequencing and molecular recognition, where molecular identity is inferred from time-dependent signals generated as individual molecules interact with a localized sensing region. For example, nanopore sequencing decodes biopolymers like DNA/RNA/peptide from dynamic signal traces as polymers pass through nanopores(*27–30*). Similarly, Raman optical strategies seek to identify biomolecules from localized, time-resolved responses(*4*, *9*, *31*, *32*), However, unlike nanopore translocation, SM-SERS hotspot sampling is stochastic and recurrent, with molecular segments randomly entering, leaving, and re-entering the enhanced field. This has so far limited direct biopolymers sequence readout from Raman trajectories(*4*, *7*, *10*). Nevertheless, sequence-related information should become extractable when these local stochastic molecular events are interpreted using an appropriate statistical framework.

In this work, CAMIL mines chain embedded nucleotide- and dinucleotide-dominated spectra directly from heterogeneous informative time-resolved DNA SM-SERS and enables the construction of a single-molecule DNA-segment spectral states library. Using this CAMIL-derived library, 12-class tri-channel multi-kernel CNN classifier maps target DNA SM-SERS trajectories into high-confidence, DNA-segment spectral state-resolved trajectories. These trajectories allow plasmonic-nanocavity-confined single molecule DNA dynamics to be statistically characterized through DNA-segment state composition, dwell length, entropy, switching frequency and transition matrix. By further developing a transition-guided scoring algorithm, state-transition matrices are converted into state-transition-edge sequence evidence, allowing all length-constrained candidate sequences to be ranked according to their probability-weighted consistency with the observed transitions and enabling reconstruction of asymmetric DNA sequences (Fig. 1F). Together, this physics-aligned deep-learning based framework integrates CAMIL-based spectral mining, CNN-based DNA segments state classification, and transition-guided dynamic sequence reconstruction, turning intrinsic SM-SERS heterogeneity into a quantitative source of single-molecular information and advancing plasmonic nanocavity SERS sensing from spectral fingerprinting toward dynamic single-molecule decoding.

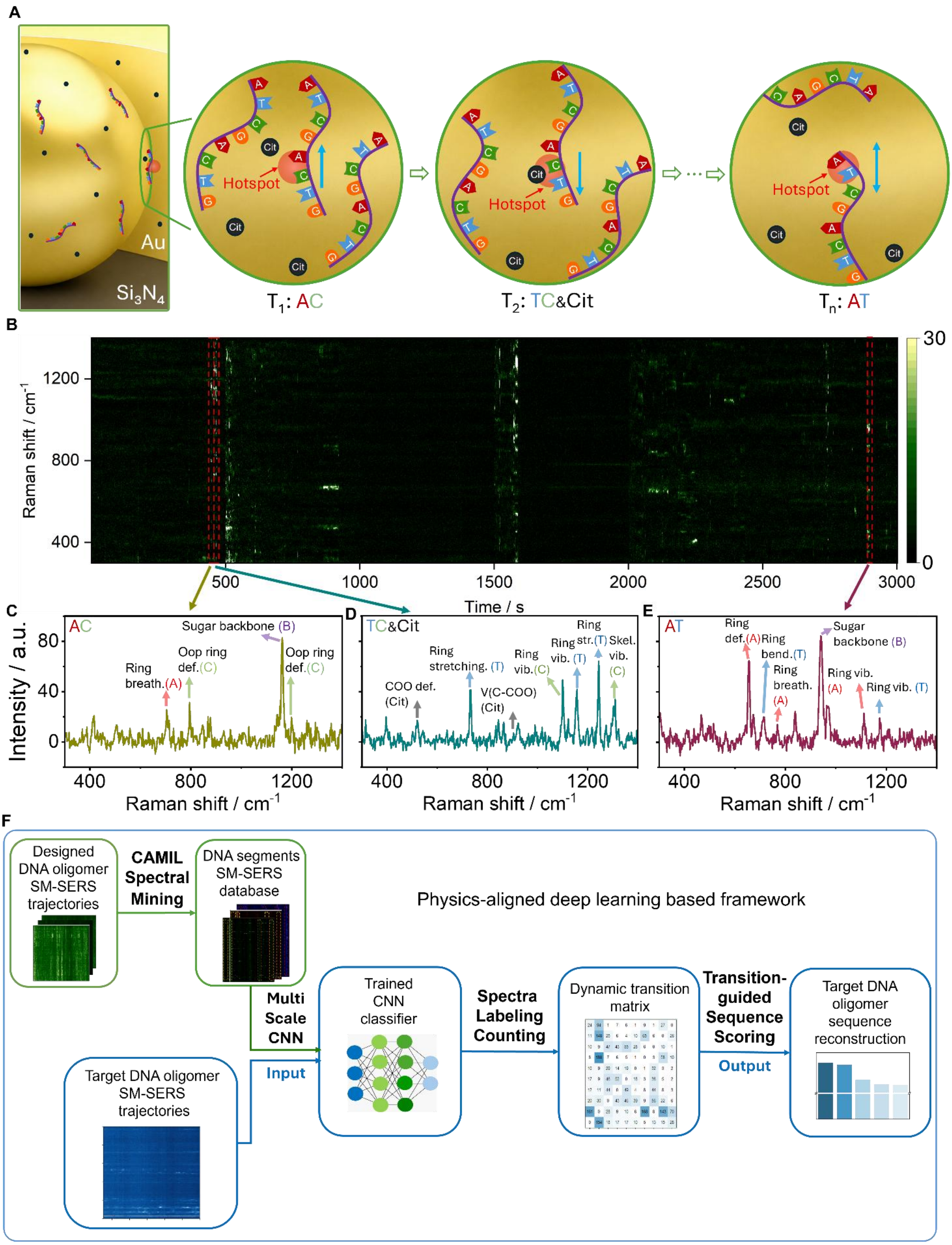

**Fig. 1. Plasmonic nanocavity-based SM-SERS heterogeneity and the physics-aligned deep learning framework for dynamic DNA oligomer decoding. (A)** Schematic of the plasmonic nanocavity based PIW platform, where an Au nanoparticle trapped within an Au nanowell forms a nanoscale hotspot at the particle–wall junction. The enlarged snapshots illustrate stochastic local readouts from different chain-embedded DNA

segments. **(B)** Time-resolved SM-SERS map of the ATCGACTG, showing strong frame-to-frame spectral heterogeneity. **(C–E)** Representative single-frame spectra from the same trajectory, corresponding to different localized DNA-segment spectral states, including AC-, TC&Cit-, and AT-dominated state contributions. **(F)** Overview of the physics-aligned deep-learning workflow, details are shown in SI 3 and 7, Figure S2 and Algorithm S1-3.

## RESULTS

### Mining Chain-Embedded DNA-segments Spectral States from Heterogeneous SM-SERS Trajectories

We first sought to clarify the physical origins of spectral heterogeneity observed in SM-SERS measurements of DNA oligomers. In the plasmonic nanocavity based PIW platform, a gold nanoparticle is trapped against the wall of a gold nanowell, forming a nanoscale hotspot at the particle–wall junction with an effective sampling region of 1~2 DNA bases (Fig. 1A)(*7*). Small molecules or local segments of larger molecules entering this highly confined hotspot are locally enhanced and detected as SM-SERS signals(*11*, *31*). The PIW substrates were prepared by assembling DNA-loaded Au nanoparticles into Au nanowells on a silicon substrate through capillary-assisted particle assembly (CAPA) method, followed by substrate drying(*7*, *33*, *34*). Consequently, citrate residues originating from the citrate-stabilized colloidal solution remain on the Au nanoparticle and nanowell surfaces, where they may contribute additional spectral components to the measured SM-SERS trajectories (Details are provided in SI S1). Beyond citrate-related signals, DNA itself introduces substantial spectral complexity. The A-, T-, C-, and G-associated SM-SERS signatures observed from DNA oligomer chains are not expected to be identical to spectra from isolated nucleobase compounds(*4*, *31*). Within an oligomer, each base can interact with neighboring bases, the phosphate–sugar backbone, the Au surface, and the local electromagnetic field in a configuration-dependent manner. Thus, the measured spectra report local base–hotspot configurations rather than isolated molecular identities alone. Representative nucleotide- and dinucleotide-dominated SM-SERS signatures must therefore be mined directly from DNA-chain trajectories instead of being assigned solely from isolated-base references.

These considerations motivated us to establish a representative SM-SERS DNA-segments spectral-state library of chain-embedded nucleotide and dinucleotide. The library was designed to capture the principal spectral states accessible within the plasmonic nanocavity hotspot, including nucleotide-, dinucleotide-, citrate-dominated, and background states. because the hotspot interrogates only a nanoscale local region of a DNA oligomer, library sequences should enrich well-defined nucleotide- and dinucleotide-level signatures while minimizing strand-design-induced structural artifacts. We therefore avoided short strands composed exclusively of C and G, such as 5′-CGCGCGC-3′, because excessively high GC content can increase duplex stability and promote hairpin formation, self-dimerization, or transient local duplex-like structures(*35*–*37*). These effects would complicate the interpretation of DNA SM-SERS trajectories by convoluting intrinsic local base–hotspot fluctuations with sequence-induced self-association or local base-pairing events.

**Table 1. Positive–negative bag design for CAMIL mining of chain-embedded nucleotide and dinucleotide signatures**

| Target | Positive bags | Segments in positive bag | Negative bags/ pre-mined negative classes | Segments in negative bag | Overlapping DNA-segments |
|---|---|---|---|---|---|
| AA | AAAAAAA | A, AA, Cit, BK | ACACACA, ATATATA, AGAGAGA | A, C, G, T, AC, AT, AG, Cit, BK | A, Cit, BK |
| CC | ACCCCCA | A, C, AC, CC, Cit, BK | ACACACA, TCTCTCT, ACGCGCA | A, C, G, T, AC, CG, TC, Cit, BK | A, C, AC, Cit, BK |
| GG | AGGGGGA | A, G, AG, GG, Cit, BK | AGAGAGA, TGTGTGT, ACGCGCA | A, C, G, T, AC, AG, CG, TG, Cit, BK | A, G, AG, Cit, BK |
| TT | ATTTTTA | A, T, AT, TT, Cit, BK | ATATATA, TCTCTCT, TGTGTGT | A, C, G, T, AT, TC, TG, Cit, BK | A, T, AT, Cit, BK |
| AC | ACACACA | A, C, AC, Cit, BK | AAAAAAA, AGAGAGA, ATATATA, TCTCTCT | A, C, G, T, AA, AG, AT, TC, Cit, BK | A, C, Cit, BK |
| AG | AGAGAGA | A, G, AG, Cit, BK | AAAAAAA, ACACACA, ATATATA, TGTGTGT | A, C, G, T, AA, AC, AT, TG, Cit, BK | A, G, Cit, BK |
| AT | ATATATA | A, T, AT, Cit, BK | AAAAAAA, ACACACA, AGAGAGA, TGTGTGT | A, C, G, T, AA, AC, AG, TG, Cit, BK | A, T, Cit, BK |
| CG | ACGCGCA | A, C, G, AC, CG, Cit, BK | AAAAAAA, ACACACA, AGAGAGA, TGTGTGT | A, C, G, T, AA, AC, AG, TG, Cit, BK | A, C, G, AC, Cit, BK |
| TC | TCTCTCT | C, T, TC, Cit, BK | ATATATA, ACACACA, TGTGTGT, ACGCGCA | A, C, G, T, AT, AC, CG, TG, Cit, BK | C, T, Cit, BK |
| TG | TGTGTGT | G, T, TG, Cit, BK | ATATATA, ACACACA, TGTGTGT, ACGCGCA | A, C, G, T, AT, AC, CG, TG, Cit, BK | G, T, Cit, BK |
| A | AAAAAAA, ATATATA | A, T, AA, AT, Cit, BK | TGTGTGT, TCTCTCT, AA, AT | C, G, T, AA, AT, CG, TC, Cit, BK | T, AA, AT, Cit, BK |
| C | ACCCCCA, TCTCTCT | A, C, T, AC, CC, TC, Cit, BK | TGTGTGT, ATATATA, AC, CC, TC | A, G, T, AT, AC, CC, TC, TG, Cit, BK | A, T, AC, CC, TC, Cit, BK |
| G | AGGGGGA, TGTGTGT | A, G, T, AG, GG, TG, Cit, BK | TCTCTCT, ACACACA, AG, GG, TG | A, C, T, AC, AG, GG, TC, TG, Cit, BK | A, T, AG, GG, TG, Cit, BK |
| T | ATTTTTA, TGTGTGT | A, G, T, AT, TG, TT, Cit, BK | ACACACA, AGAGAGA, AT, TG, TT | A, C, G, AC, AG, AT, TG, TT, Cit, BK | A, G, AT, TG, TT, Cit, BK |

Note: DNA-segment labels in the negative-bag column, such as AA, AT, AC, AG, TG, and TT, indicate previously mined target-dominated spectral classes, not additional measured DNA oligomer sequences.

On this basis, we selected ten model DNA oligomers for DNA-segments spectral-state library construction: AAAAAAA, ACCCCCA, ATTTTTA, AGGGGGA, ACACACA, AGAGAGA, ATATATA, TCTCTCT, TGTGTGT, and ACGCGCA. These sequences were designed as controlled reference sets enriched in defined local nucleotide and dinucleotide patterns for mining representative chain-embedded DNA-segment SM-SERS signatures directly from DNA oligomer trajectories. AAAAAAA served as an A- and AA-dominated reference oligomer. ACCCCCA, ATTTTTA, and AGGGGGA were designed as A-flanked C-/CC-, T-/TT-, and G-/GG-enriched oligomers to promote DNA-segment dominated spectral states while avoiding fully C- or G-only constructs. ACACACA, AGAGAGA, ATATATA, TCTCTCT, and TGTGTGT enriched repeated AC, AG, AT, TC, and TG local dinucleotide patterns, whereas ACGCGCA provided a moderately CG-rich, A-flanked oligomer that preserves local CG patterns while avoiding extreme 100% GC content (Table 1). Together, this sequence panel establishes a controlled experimental basis for constructing a SM-SERS DNA-segments spectral-state library, physically aligned with the confined sampling regime of the hotspot and supports the downstream interpretation of dynamic DNA SM-SERS trajectories from other more complex DNA oligomers.

**CAMIL-Mined Spectral States Reveal SM-SERS Chain-Embedded DNA-segment Signatures**

Following this rationale, we developed a CAMIL-based workflow to mine citrate- and chain embedded nucleotide-/dinucleotide-dominated spectra from heterogeneous SM-SERS trajectories. First, all raw time-resolved SM-SERS datasets were processed using the same data-preprocessing workflow, including baseline correction, quality filtering, and signal selection(SI S2 and Table S1-S3)(*38*). For each analyte, average spectra of signal-containing frames and peak-occurrence event histograms were generated to summarize recurrent vibrational features across dynamic trajectories (Fig. 2E, 2F, Fig. S3, S7, S10, S13, S16, S19, S21, S23, S25, and S27)(*7*, *32*). Although useful as an overview, these analyses average over heterogeneous contributions from citrate, chain-embedded nucleotide and dinucleotide patterns, backbone vibrations, background and mixed local state signals, therefore could not resolve individual spectra corresponding to target-dominated states.

CAMIL was first used to analyze citrate-only PIW samples and establish citrate-related spectral states. Citrate-containing trajectories spectra were used to construct positive bags, whereas signal-poor background spectra served as contrastive negative bags, enabling enrichment of citrate-dominated SM-SERS signatures. Specifically, different normalization methods in CAMIL were found substantially affect the mined citrate-related features: raw-intensity inputs highlighted citrate-discriminative spectra mainly in the 400–900 $cm^{-1}$ region, whereas normalized inputs emphasized complementary relative spectral profiles, including recurrent $COO^-$- and $CH_2$-related features (Figure S4 and Table S6). The overlap among high-ranking spectra obtained under different normalization strategies supports the reliability and robustness of CAMIL and indicates that citrate contributes multiple representative spectral states within heterogeneous SM-SERS trajectories. Details are provided in Figs. S4-S6 and Table S6.

Based on these results, CAMIL was then applied to the ten designed DNA model oligomers with three complementary inputs: raw-intensity, z-score-normalized, and min–max-normalized spectra. Taking AC as an example, ACACACA-derived positive bags contained AC-dominated spectra together with A-dominated, C-dominated, citrate-dominated, backbone-associated, background, and mixed local states, whereas ATATATA and TCTCTCT were used as negative bags because they retain overlapping A-, C-, backbone-, and citrate-related contributions but lack the local AC pattern (Fig. 2A and B). During the CAMIL processing, within the positive bags, the trained context-averaged attention score reflects how strongly each spectrum represents the target local state. Spectra with clearer AC-dominant features receive higher scores, whereas spectra dominated by non-target contributions, such as A-, C-, citrate-dominated, background, or mixed states, receive lower scores (Fig. 2C). High-scoring spectra from ACACACA-derived positive bags were therefore selected as AC-dominated representative spectra (Fig. 2D, additional details in S3.1-3.10, Algorithm S1). The representative AC-dominated spectra mined from raw-intensity, z-score-normalized, and min–max-normalized inputs are shown in Fig. 2G–I. and their peak-occurrence profiles were integrated to obtain a combined AC-dominated spectral feature profile (Fig. 2J–M). representative set showed recurrent bands at 400–430, 590–620, 700–730, 780–810, 930–960, 1140–1170, 1190–1220, and 1240–1270 $cm^{-1}$, which can be tentatively assigned to adenine-, cytosine-, and backbone-associated vibrations (Fig. 2I and Table S15)(*31*, *39*–*41*). The coexistence of tentatively assigned A-, C-, and backbone-related features supports the assignment as AC-dominated local DNA-segment states. Representative SM-SERS spectra and tentative peak assignments for the remaining nucleotide- and dinucleotide-dominated classes, including A, C, G, T, AA, CC, GG, TT, AG, AT, CG, CT, and TG, are summarized in Figs. S7–S28 and Tables S7–S20. All high-score spectra obtained across the CAMIL were combined into class-specific representative spectrum sets. This spectrum-set representation preserves both the shared vibrational identity of each target class and the intrinsic spectrum-to-spectrum variability arising from hotspot sampling of different DNA-segment configurations. Thus, the single molecule DNA-segment spectral-state library was built from data-mined, chain-embedded SM-SERS DNA trajectory spectra rather than isolated compounds or ensemble-averaged references. Together, these results demonstrate that CAMIL can extract representative nucleotide- and dinucleotide-dominated DNA-segment spectra directly from heterogeneous SM-SERS trajectories, providing a specific SM-SERS DNA-segment spectral-state library for downstream classification and trajectory interpretation.

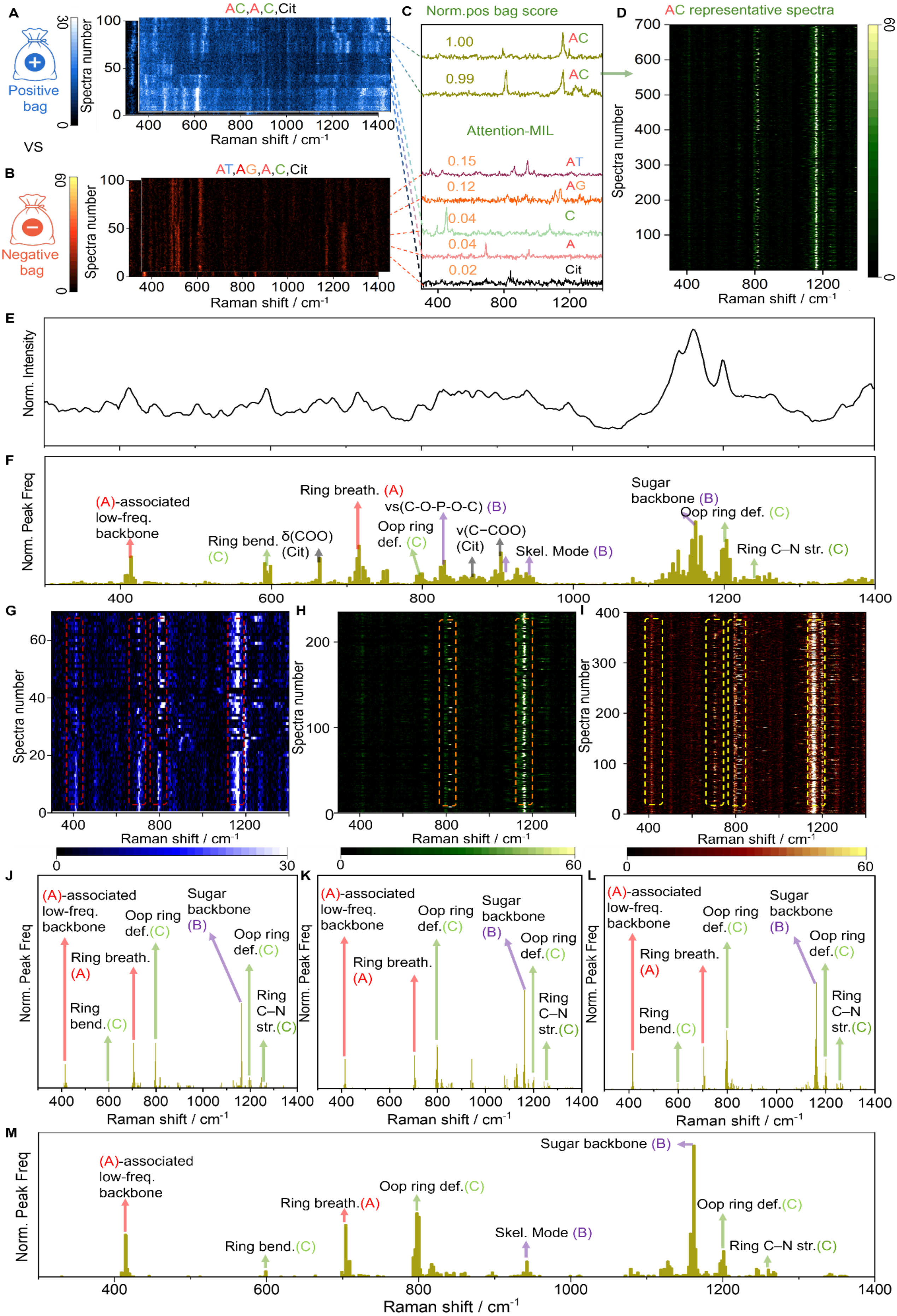

A
Positive bag
VS
B
Negative bag
AC,A,C,Cit
AT,AG,A,C,Cit
Spectra number
Raman shift / cm-1
C
Norm.pos bag score
1.00
0.99
AC
Attention-MIL
0.15
0.12
0.04
0.04
0.02
AT
AG
C
A
Cit
D
AC representative spectra
E
Norm. Intensity
F
Norm. Peak Freq
(A)-associated low-freq. backbone
Ring bend. (C)
δ(COO) (Cit)
Ring breath. (A)
vs(C-O-P-O-C) (B)
Oop ring def. (C)
v(C−COO) (Cit)
Skel. Mode (B)
Sugar backbone (B)
Oop ring def. (C)
Ring C–N str. (C)
G
H
I
J
K
L
M
Ring breath.(A)
Ring bend.(C)
Oop ring def.(C)
Skel. Mode (B)
Sugar backbone (B)
Ring C–N str.(C)

**Fig. 2. Data-mined AC-dominated representative spectra from ACACACA. (A, B)** CAMIL strategy for extracting AC-dominated representative spectra, with ACACACA spectra used as positive bags and spectra from structurally related oligomers lacking the AC pattern, including ATATATA and TCTCTCT, used as negative bags. **(C)** CAMIL suppresses shared spectral contributions while enriching AC- dominated features, enabling attention-based scoring of spectra within positive bags. **(D)** High-score spectra selected from the positive bags represent AC-dominated SM-SERS signatures for spectral-state library construction. **(E)** Averaged spectrum and **(F)** peak-occurrence event histogram of signal-containing ACACACA frames selected using the workflow in SI S2. **(G–I)** AC-dominated spectra mined using raw, z-score, and min–max inputs; dashed boxes mark recurrent regions. **(J–L)** Corresponding peak-occurrence histograms from (G–I). **(M)** Combined histogram summarizing AC-dominated features across 3 input strategies of CAMIL. Tentative assignments in Table S15.

This library further reveals several key features of the SM-SERS DNA-segment states spectra. First, the data-mined representative SM-SERs DNA-segment states spectra do not necessarily reproduce corresponding isolated nucleobase references. For example, the representative A- and AA-dominated SM-SERS spectra show the dominant recurrent feature near 690–700 cm$^{-1}$, rather than the ~730 cm$^{-1}$ region commonly observed for isolated adenine. This observation is consistent with our previous particle-in-nanopore single-molecule results and related SM-SERS observations(*4*, *42*, *43*), indicating that single-molecule vibrational signatures can be reshaped by adsorption geometry, local electromagnetic coupling, and mode-selective enhancement(*15*, *44*).

Second, several recurrent spectral regions, particularly in the low-frequency region of 300–500 cm$^{-1}$, cannot be assigned unambiguously using conventional isolated-base references(*39*, *40*). These bands likely arise from mixed contributions of ring deformation, backbone motion, citrate deformation, and metal–molecule coupling rather than from a single localized bond vibration. Because these modes are highly sensitive to molecular orientation, adsorption geometry, and hotspot coupling, weak or surface-activated vibrations can become selectively enhanced under specific configurations. In addition, SM-SERS features from A, C, T, G, and the sugar–phosphate backbone frequently overlap, indicating that peak assignment in SM-SERS cannot be treated as a simple one-to-one mapping to standard reference tables.

Third, dinucleotide-dominated representative spectra are not simple linear combinations of their corresponding single-base spectrum. Instead, they retain partial similarity to their component bases while exhibiting specific peak patterns and selective enhancement behavior. For example, adenine-related features vary across different A-containing dinucleotide: the A-associated band at 710–740 cm$^{-1}$ is more prominent in AC-dominating states, whereas AG-dominating states show additional A-associated features around 670–700 and 1340–1370 cm$^{-1}$ (fig. S20 and table S16). This behavior indicates that neighboring-base interactions, local stacking, backbone connectivity, gold-surface adsorption, and local field coupling jointly shape the observed SM-SERS DNA-segment fingerprints.

Notably, the representative spectra of single nucleotide- and their corresponding homo-dinucleotide exhibited overall consistency while minor differences in relative peak contributions (Fig S8&S9, S11&S12, S14&S15, and S17&S18). For example, the A- and AA-SM-SERS spectrum sets share adenine-associated vibrational features, but the adenine contribution was more pronounced in AA-dominated spectra to the sugar–phosphate backbone, whereas A-dominated spectra retain more visible backbone-related features. Similar relationships were observed for the C/CC-, G/GG-, and T/TT-dominated classes, where the corresponding nucleotide- and homo-dinucleotide-dominated spectra exhibited stronger mutual similarity. This trend is physically reasonable because the mined A-, C-, G-, and T-dominated spectra represent base-dominated local states within DNA chains rather than isolated bases. Each nucleobase remains covalently connected to the sugar–phosphate backbone, and the confined hotspot samples a local chain segment of 1~2 DNA base rather than a chemically isolated base. Backbone-associated vibrations can therefore naturally coexist with base-dominated features in the mined SM-SERS DNA-segments states. The base-dependent similarity differences of nucleotide- and homo-dinucleotide-dominated spectra may further reflect different adsorption affinities on Au: strong A–Au affinity makes AA-state spectra dominated by the adenine-associated modes, whereas weaker C-, G-, and T–Au interactions may allow backbone contributions to remain more visible, leading to greater C/CC, G/GG, and T/TT spectral similarity(*4*, *45*–*47*).

**Tri-Channel CNN Classification and Probability-Guided DNA-segment State Assignment**

To better exploit the DNA-segment spectral-state library mined by CAMIL, we fed them into a tailored tri-channel multi-kernel CNN. The spectra were represented over the Raman-shift range of 300–1400 $cm^{-1}$, with each spectrum represented containing 456 intensity points, and were randomly split into training, validation, and test sets at a ratio of 0.70/0.15/0.15. For each representative spectrum, three complementary input channels were constructed from the raw spectrum, the z-score-normalized spectrum, and the min–max-normalized spectrum. This design allowed the classifier to jointly learn absolute-intensity information, standardized spectral profiles, and relative peak-distribution patterns. The tri-channel input tensor was first processed by parallel shallow convolutional branches for initial feature extraction and feature fusion. The fused feature maps were then passed through multi-kernel one-dimensional convolutional branches with different kernel sizes to capture Raman features across multiple spectral scales, including narrow peaks, broader bands, and local peak patterns. After branch concatenation, residual convolutional blocks, global average pooling, and a LayerNorm–MLP classification head were used to generate class logits and final predictions. Additional details are shown in SI S3.11-3.17, Fig S2, and Algorithm S2.

With the CAMIL–3CNN workflow established, we first attempted a 16-class classification task, in which nucleotide (A, C, G, T), homo-dinucleotide (AA, CC, GG, TT), hetero-dinucleotide (AC, AG, AT, CG, TC, TG), citrate (Cit), and background (BK) were treated as independent classes (Table S21). However, this setting showed limited performance because the nucleotide and corresponding homo-dinucleotide were highly similar (Fig S29). This result is consistent with the spectral analysis above: A/AA, C/CC, G/GG, and T/TT share substantial base- and backbone-associated features because the hotspot sampling region is sufficiently small to probe chain-embedded local states, yet large enough to capture coupled base–backbone contributions rather than chemically isolated molecular units. Thus, forcing these closely related spectra into separate labels introduced artificial ambiguity beyond the effective molecular resolution of the plasmonic nanocavity hotspot readout.

We therefore merged each nucleotide/homo-dinucleotide pair into a broader base-dominated category, yielding A-, C-, G-, and T-dominated classes and a physically better-aligned 12-class classification task (Table S22). This label structure preserves the distinction among nucleotide-, hetero-dinucleotide, Cit, and BK states, while avoiding over-separation of spectrally similar nucleotide and homo-dinucleotide states. The resulting 12-class tri-channel multi-kernel CNN showed strong classification performance. To minimize the influence of unequal spectrum numbers among classes, balancing and evaluation strategies were applied as described in SI S3.16-3,17, Fig S2, and Algorithm S2. The row-normalized confusion matrix showed dominant diagonal responses across all classes, yielding an average accuracy of 86.2±0.6% over 5 independent runs and indicating that most spectra were correctly assigned with limited cross-class leakage (Table S24, Fig. 3A). One-vs-rest ROC analysis further showed high class separability across the 12 classes (Fig. 3B), and the precision–recall analysis confirmed generally balanced class-wise performance (Fig. 3C). More classification performance details are shown in Table S23. These results indicate that CAMIL-mined representative DNA-segment spectra contain sufficient discriminative information for downstream classification when the label structure is aligned with the physical resolution of plasmonic nanocavity-based SM-SERS.

To benchmark the proposed CAMIL–3CNN strategy, we compared it with several conventional classifiers, including K-nearest neighbors (KNN), artificial neural network (ANN), convolutional neural network (CNN), random forest (RF), and support vector machine (SVM) models using randomly sampled spectra matched in class-wise sample size to the CAMIL-3CNN dataset (Details are shown in SI S4, Table22). Their confusion matrices are shown in Figs. S30 to S35. These classifiers achieved lower accuracies of 77.3%, 48.5%, 80.5%, 71.5%, and 65.9%, respectively. Their reduced performance mainly reflects the intrinsic spectral heterogeneity of the original DNA SM-ERS trajectory. Although the ten DNA oligomers are sequence-distinct, randomly sampled spectra from each DNA SM-ERS trajectory contain overlapping components from citrate background, common backbone contributions, and shared nucleotide and dinucleotide contributions. These nonspecific or shared spectral components blur the class boundaries and introduce conflicting label information, causing classifiers was trained directly on random samples of trajectory-level labels to learn mixed or weakly informative features. Importantly, accuracy alone does not fully address the central objective of this study. Even when conventional classifiers achieved relatively high accuracy, such as KNN with accuracy of 80.5%, they remained limited in mechanistic interpretability because they were optimized for label prediction rather than segment-aware spectrum mining or chemical-state analysis. They do not explicitly identify which spectra carry target DNA-segment-state information, distinguish DNA-segment-dominated spectra from citrate- or background-dominated signals, or explain how nanocavity-confined molecular configurations contribute to the observed SM-SERS heterogeneity. In contrast, the CAMIL–3CNN workflow first enriches DNA-segment-relevant spectra through contrastive attention-based mining and then performs classification with a physically aligned label structure. This strategy reduces spectral ambiguity before classification, improves predictive performance, and links SM-SERS heterogeneity to interpretable local DNA-segment-state assignments.

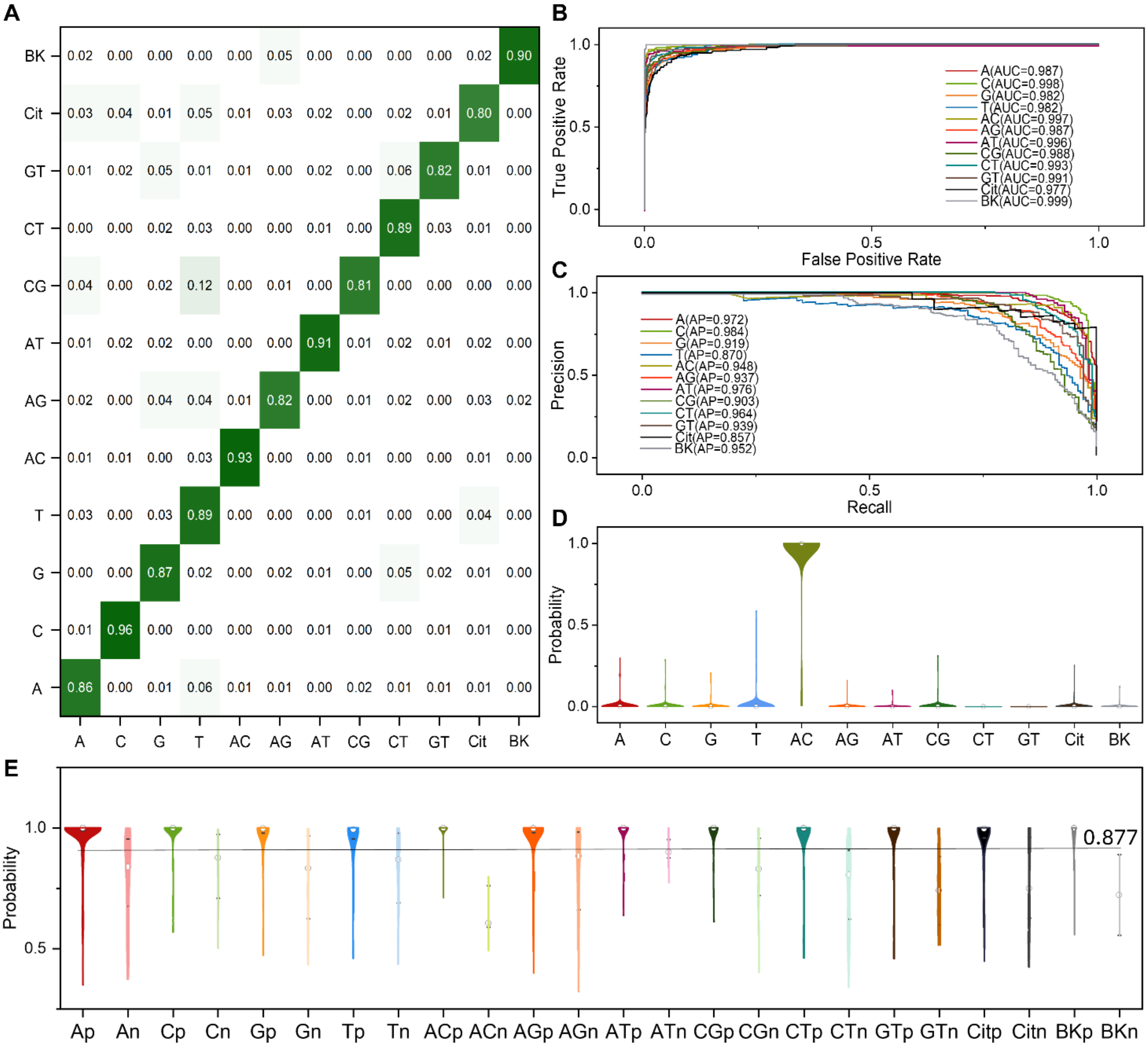

**Fig. 3. Classification performance and probability analysis of the tri-channel multi-kernel CNN. (A)** Row-normalized confusion matrix for the 12-class classification task, showing the prediction distribution for each true class. **(B)** One-vs-rest ROC curves and corresponding AUC values for all 12 classes, evaluating class-wise discrimination performance. **(C)** Precision and recall of the 12 classes, summarizing class-specific classification sensitivity and reliability. **(D)** AC-class output probabilities for all spectra predicted as AC, including both true AC spectra and false-positive assignments from other classes. **(E)** Output probability distributions for positive and negative recognition events for each class. Class colors are kept consistent throughout this work.

To evaluate model reliability beyond final class labels, we analyzed the output probability distributions of a trained CAMIL-3CNN model. Using AC as an example, we examined AC-class output probabilities for all spectra predicted as AC, including true AC spectra and false-positive assignments from other classes (Fig. 3D). False-positive AC assignments generally showed probabilities below 0.65, indicating that these spectra were ambiguous even though AC was the highest-probability class. In contrast, correctly classified AC spectra typically showed AC probabilities above 0.8, consistent with high-confidence recognition of AC-dominated spectral states. We further compared positive and negative recognition events across all classes. (Fig. 3E). For example, A-positive (Ap) denotes spectra truly belonging to the A-dominated class and correctly predicted as A, whereas A-negative (An) denotes non-A-dominated spectra incorrectly predicted as A because their A-class probability was still the highest. Across the 12 classes, correct positive assignments generally showed higher and more concentrated probabilities, with distribution centers around 0.941. whereas false-positive assignments showed lower and broader probability distributions, with centers below 0.877. These results indicate that most incorrect cross-class assignments occurred under lower-confidence conditions, while correctly recognized spectra were assigned with strong model confidence. Together, these probability-level analyses demonstrate that the 12-class CAMIL–3CNN model provides not only accurate classification but also interpretable confidence behavior, helping to distinguish robust DNA-segment state assignments from ambiguous cross-class predictions.

**Trajectory-Level Transition Analysis for SM-SERS DNA Decoding**

We therefore adopted a statistical trajectory-level strategy based on the trained CAMIL–3CNN classifier. A maximum output probability of 0.9 was used as a confidence threshold: spectra above this threshold were assigned to their predicted DNA-segment states, whereas lower-confidence spectra were labeled as uncertain. Using this probability-filtered framework, the targeted time-resolved SM-SERS DNA trajectory was converted into a high-confidence DNA-segment states sequence. This representation enabled relative, confidence-filtered quantitative analysis of its single-molecule DNA dynamics, including segment states composition, dwell lengths, entropy, switching frequency, and transition matrix. Together, these descriptors provide a statistical view of how local DNA segment states of query DNA oligomer appear, disappear, and interconvert within the hotspot, thereby linking heterogeneous SM-SERS DNA trajectories to dynamic DNA-segment states behavior.

Under a stringent probability threshold of 0.9, the ATCGACTG SM-SERS trajectories showed a non-uniform molecular-state distribution across approximately 33,000 spectra from 70 trajectory files (Fig. S35; details in SI S5.1). This indicates that plasmonic nanocavity hotspot does not sample all local DNA segment states equally but preferentially detects specific nucleotide- and dinucleotide-dominated states depending on local adsorption, accessibility, and SERS enhancement. Component-level dwell length, which describes confidence-filtered apparent persistence rather than absolute molecular residence time, showed that most molecular-state assignments appeared as short consecutive runs (Fig. S36A–K; details in SI S5.2). While molecule-level dwell length, in which all DNA-related labels were grouped as signal states, gave longer apparent dwell length because one DNA-signal period can contain multiple local state assignments (Fig. S36 and Table S25; details in SI S5.2–S5.3). Global trajectory behavior was further summarized using three descriptors: state entropy H, molecule-level dwell length T, and switch rate R. These HTR descriptors captured state diversity, DNA-signal persistence, and transition frequency of DNA oligomers, which could reproducibly separate ATCGACTG, AAAAAAA, and TACAAGTAAAG after trajectory-file-level resampling (Fig. S37-S38; details in SI S5.4–5.5, S6).

We further constructed transition matrix from adjacent high-confidence state assignments, excluding uncertain labels and self-transitions (SI S5.6). Although the CAMIL-3CNN classifier achieved an overall accuracy of approximately 86.2%, and a stringent probability threshold of 0.9 was applied to retain high-confidence molecular-state assignments, misclassifications and missed assignments cannot be completely excluded. Nevertheless, within the same probability-filtered transition matrix, clear local DNA-segment patterns remained. The schematic local motion and its transition matrix representation are illustrated in Fig. 4A–C. Fig. 4A shows a representative time-resolved ATCGACTG SM-SERS trajectory. Because each spectrum was acquired with an integration time of 0.1 s, the time indices on the vertical axis correspond to real time after multiplication by 0.1. Fig. 4B shows the spectra and their predicted DNA-segment-state labels within the 3.1–4s time window by our trained CAMIL-3CNN model. These time-resolved spectra were intentionally selected because it contains relatively few uncertain assignments while showing clear state-transition events. Within this one-second window, the ATCGACTG SM-SERS trajectory first shows a citrate-dominated state, followed by the emergence of AC-dominated signals. The dominant state then switches to T, persists for one frame, and subsequently changes to TC, C, and again T, before reaching an uncertain assignment and a background-like state. This frame-by-frame evolution illustrates how different local DNA-segment states in ATCGACTG oligomer enter and leave the hotspot over time. Fig. 4C schematically depicts the corresponding back-and-forth motion of the ATCGACTG molecule within the hotspot. The insert shows a simplified transition matrix extracted from this short time trajectory spectra, summarizing the observed state-to-state transitions. In this way, the dynamic SM-SERS trajectory is transformed into a mathematical representation of local segment states motion within the hotspot. For example, in ATCGACTG (Fig. 4D), frequent A-to-C transitions were observed for A and C are adjacent in the sequence. Similarly, the high transition frequency from C to AT is consistent with continuous sampling of locally accessible sequence-related DNA segment states within the hotspot. In contrast, CG and TG are not directly adjacent and are separated by intervening AC segments. Accordingly, only one CG-to-TG transition was observed, which may arise from model misclassification, threshold-induced omission of intermediate states, or some molecular motion occurring faster than the single-frame acquisition time. The contrast between frequent adjacent-state transitions and rare distant-segment transitions supports a physical picture in which DNA molecules mainly undergo local, directionally reversible back-and-forth surface motion rather than long-range random jumping, which also consistent with the physical picture proposed in our previous work(*4*, *7*, *10*). Thus, the transition matrix provides a more detailed view of how local DNA segment motion is encoded in plasmonic nanocavity-based DNA SM-SERS trajectories.

Based on this representation, we next investigated whether DNA sequence information could be inferred from the transition matrix. In our system, DNA molecules do not pass through the plasmonic hotspot along a simple unidirectional path. Instead, they undergo stochastic back-and-forth motion within the hotspot. The resulting data therefore differs fundamentally from conventional sequencing reads, which typically rely on (*48*, *49*). Consequently, overlap-based de novo assembly, dynamic programming, and hidden Markov modeling are not well matched to this measurement regime(*50*, *51*). In addition, stochastic back-and-forth motion allows the same local sequence segment to be sampled repeatedly in either the forward or reverse direction. Highly symmetric sequences are therefore difficult to distinguish and reconstruct reliably, as shown in Fig. S39. Sequence reconstruction was thus mainly applied to asymmetric sequences.

Because the transition matrix provides a physically meaningful statistical representation of how local molecular states of asymmetric sequences are repeatedly sampled and interconverted, we developed a transition-guided sequence scoring algorithm to infer sequence information from these transition matrices (Fig. 2C; details in SI S7, Table S27 and Algorithm S3). In this algorithm, transition edges between valid nucleotide and dinucleotide DNA-segment states were first converted into compatible local segment evidence. Then, for a target sequence length $L$, all candidate sequences were exhaustively enumerated. Each candidate was scored according to how well it explained the transition-derived segment evidence. The highest-ranked asymmetric candidate was then identified as the sequence that best reconciled the observed local transition statistics. This strategy is well suited to sparse, short-range, and directionally ambiguous SM-SERS observations, which uses the transition matrix as a statistical representation of local sequence relationships, making the framework consistent with the intrinsic dynamic heterogeneity of SM-SERS.

We then applied the transition matrix of ATCGACTG from Fig. 4D to the transition-guided constrained sequence scoring algorithm. Ranked candidate sequences were generated for different assumed sequence, including $L$ = 8, $L$ = 7, and $L$ = 9 (Fig. 4F–H). When the correct length was used ($L$=8), the measured sequence ATCGACTG was ranked first. When incorrect lengths were tested ($L$=7 and $L$=9), the algorithm still recovered candidates closely related to the measured sequence. For example, the top-ranked candidate at $L$ = 7 was ATCGACT, and the second-ranked candidate at $L$ = 9 was ATCGACTGA. These results indicate that the transition matrix contains sequence-related local statistical information. However, candidates generated with incorrect assumed lengths cannot fully reproduce the measured sequence. Thus, when using the trained CAMIL-3CNN model with a reasonable probability threshold of 0.9 and prior knowledge of the target sequence length, the transition matrix can support reconstruction of the measured DNA sequence from SM-SERS trajectories.

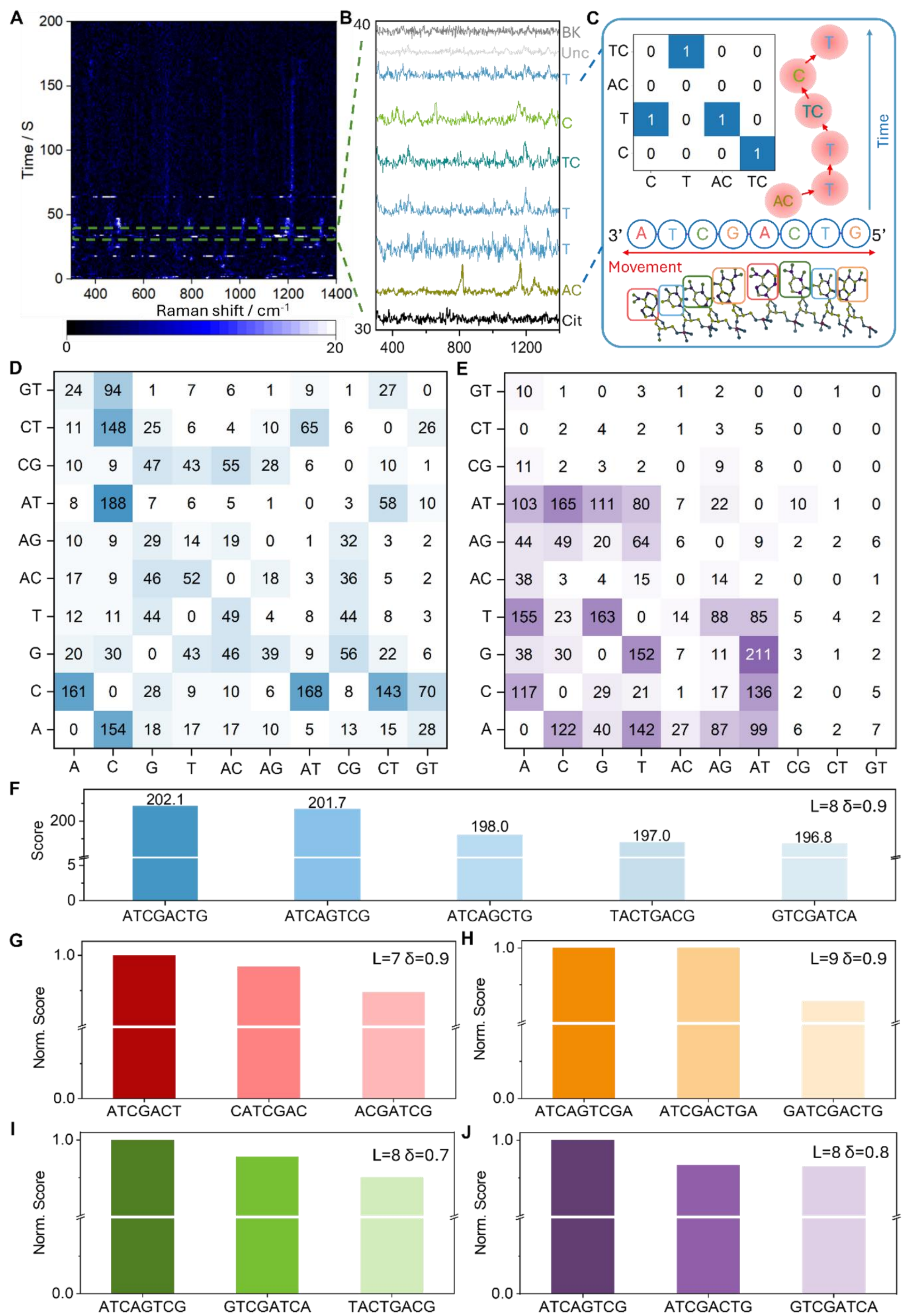

**Fig. 4. Transition-guided sequence reconstruction from transition statistics (A)** Representative time-resolved SM-SERS trajectory of the ATCGACTG. The dashed green box marks the time-window spectra selected for frame-by-frame analysis. **(B)** Classified spectra and corresponding DNA-segment state labels

from the selected 3.1–4.0 s time window. **(C)** Schematic illustration of the corresponding back-and-forth ATCGACTG oligomer motion within the hotspot and the transition matrix extracted from this selected time-resolved trajectory. **(D, E)** Transition matrices of adjacent high-confidence predicted DNA-segment labels from ATCGACTG and TACAAGTAAAG SM-SERS trajectories, respectively. **(F–H)** Ranked candidate sequences obtained from the ATCGACTG transition matrix using different assumed lengths: $L$=8, $L$=7, and $L$=9 with threshold (δ) are all 0.9. **(I–J)** Effect of confidence threshold on reconstruction at $L$=8. Lowering the δ to 0.7 reduced the rank of ATCGACTG fall outside the top three candidates, whereas at 0.8 it ranked second.

We further examined how the confidence threshold of trained CAMIL-3CNN classifier affects sequence reconstruction. Lower thresholds, such as 0.7 or 0.8, retain more assigned molecular-state labels but also introduce more false-positive assignments, thereby altering the transition-frequency distribution. When the transition matrix generated at a threshold of 0.7 was used for scoring with $L$=8 (Fig. S40), the true sequence ATCGACTG did not be ranked among the top three candidates (Fig. 4I). This result indicates that excessive false-positive assignments can distort transition statistics and reduce reconstruction accuracy. By contrast, when the threshold was increased to 0.8 (Fig. S41), ATCGACTG was ranked second for $L$=8 (Fig. 4J). The comparison among Fig. 6F, I, and J further highlights the importance of the confidence threshold selection. When false-positive assignments are reduced and high-confidence spectral labels dominate the transition matrix, the resulting transition statistics more faithfully preserve the underlying sequence structure, leading to improved candidate-sequence ranking.

Although stochastic back-and-forth motion within the hotspot makes highly symmetric DNA sequences difficult to reconstruct directly, we further investigated whether asymmetric information could still be recovered from sequences containing symmetric units. We used TACAAGTAAAG as a representative case. Repetitive or locally symmetric segments were first condensed: ACA, AA, and AAA were simplified as A. After this reduction, the asymmetric component of the original strand was extracted as TAGTAG. Accordingly, we simplified the transition matrix derived from the TACAAGTAAAG trajectories by removing all C-containing states, yielding a reduced transition matrix shown in Fig. S42. This reduced matrix was then used as input for the transition-guided constrained sequence scoring algorithm. With a target length of $L$=6 and a probability threshold of 0.9, the reconstructed sequence TAGTAG was ranked first (Fig. S43). To further evaluate length dependence, the same reduced transition matrix was also tested with $L$ = 4, 7, and 8 (Figs. S44–S46). These alternative assumed lengths generated shorter or extended TAG-related candidates, such as TAGT, TAGTAGT, and TAGTAGTA, indicating that the reduced transition matrix consistently preserves TAG-related transition evidence. However, only the target length of $L$ = 6 recovered TAGTAG as the most direct asymmetric component of TACAAGTAAAG. These results show that appropriate physical simplification and transition-matrix reduction can recover embedded asymmetric sequence information from partially repetitive DNA trajectories, while also highlighting the dependence of reconstruction on the assumed sequence length. Overall, the transition matrix captures both the underlying DNA sequence structure and the local molecular dynamics within the hotspot. By converting stochastic molecular motion into state-to-state transition statistics, sequence-relevant information can be extracted from dynamic SM-SERS trajectories through an appropriate statistical framework.

## DISCUSSION & OUTLOOK

In this work, we establish a physics-aligned framework for extracting DNA-segments state and sequence-level information from heterogeneous SM-SERS trajectories in plasmonic nanocavity. Unlike conventional Raman or SERS measurements, where spectra are commonly interpreted as average molecular fingerprints, the SM-SERS platform records dynamic streams of local, configuration-dependent spectral molecular states. As the hotspot samples a sub-molecular region comparable to 1~2 DNA bases, each spectrum represents a transient local interaction among the DNA oligomer, the gold surface, residual citrate, and the nanoscale electromagnetic field(*4*, *7*, *10*, *15*, *18*). Spectral heterogeneity is therefore not merely experimental noise, but an intrinsic consequence of confined single-molecule sampling. By integrating CAMIL with a tri-channel multi-kernel CNN, we converted heterogeneous SM-SERS trajectories into chemically interpretable DNA-segment-state assignments. With three complementary normalization inputs, CAMIL mined nucleotide- and dinucleotide-dominated spectra directly from contrastive native DNA trajectories, enabling multi-perspective feature extraction without manual frame-level annotation or isolated-base reference spectra. The 12-class classifier was likewise physics-aligned, incorporating single-nucleotide states together with their corresponding homo-dinucleotide states to match the approximately 1~2 DNA bases sampling scale of the plasmonic hotspot. In this way, CAMIL-3CNN based spectral mining and classifier design were matched to the effective hotspot readout of the plasmonic nanocavity platform.

Trajectory-level analysis further demonstrates that measured SM-SERS contains information beyond static spectral classification. DNA segment-state composition, dwell length, switch rates, entropy, and transition matrix collectively describe how local DNA regions enter, persist within, and leave the hotspot. These descriptors reveal sequence-dependent dynamic fingerprints among different DNA oligomers. Notably, the transition matrix provides a statistical bridge between stochastic segment motion and sequence inference. In the plasmonic nanocavity hotspot, DNA oligomer does not pass through the sensing volume as a deterministic, unidirectional read. Instead, local back-and-forth motion causes adjacent or nearby sequence elements to be repeatedly sampled in both directions. This measurement physics differs fundamentally from conventional sequencing reads and explains why direct overlap-based assembly or deterministic path-reconstruction strategies are poorly suited to our plasmonic nanocavity-based SM-SERS sensing. However, the same stochastic reversibility also generates repeated local transition statistics. By converting transitions between high-confidence segment states into state-transition-edge sequence evidence, candidate sequences can be ranked according to their probability-weighted consistency with the observed transition structure. The successful reconstruction under a defined sequence length and stringent confidence threshold demonstrates that sequence information can be recovered from dynamic SM-SERS trajectories through a statistical decoding strategy.

Despite these advantages, the present framework also has important limitations. First, sequence reconstruction depends strongly on the reliability of DNA-segment state classification. False-positive state assignments can introduce artificial transition edges and distort the transition matrix, whereas overly stringent filtering may discard useful but lower-confidence frames. Thus, confidence thresholding represents a trade-off between information retention and transition reliability. Second, highly symmetric or repetitive sequences remain difficult to reconstruct because local back-and-forth sampling can produce directionally ambiguous or degenerate transition statistics, and different candidate sequences may generate similar local transition patterns, limiting the ability of the current algorithm to resolve sequence direction or repeated segments. The requirement for prior knowledge of sequence length is another practical constraint. In the present implementation, candidate-sequence scoring is performed under a predefined length constraint. Although incorrect length assumptions still provide informative local sequence evidence, indicating that transition-guided scoring preserves partial sequence information even when the assumed length is imperfect. Nevertheless, accurate reconstruction still benefits from a reasonable length constraint. Therefore, while this length-constrained strategy is suitable for proof-of-concept validation with designed oligomers, future applications will require algorithms capable of jointly inferring sequence length, directionality, and sequence composition.

Future improvements should therefore focus on both experimental and computational advances. Experimentally, improving hotspot stability, enhancing detection sensitivity, shortening the spectral acquisition time, and better controlling DNA adsorption geometry would improve the reliability of DNA-segment-state trajectories. Computationally, uncertainty-calibrated classifiers and larger trajectory libraries could further strengthen the connection between local transition statistics and sequence inference. incorporating additional physical constraints, such as strand orientation, adsorption stability, or controlled molecular motion, may further improve reconstruction accuracy.

In conclusion, this study reframes single-molecule SERS heterogeneity as an information-rich signal rather than an analytical obstacle. In confined plasmonic geometries, spectral fluctuations encode the stochastic exploration of local molecular configurations. When interpreted using physics-aligned machine learning and transition-level statistics, these fluctuations can reveal molecular-state dynamics and sequence-related structure. This strategy provides a conceptual foundation for dynamic SM-SERS-based molecular decoding, in which sequence information is inferred not from a continuous linear readout, but from the statistical organization of sparse, transient, and locally enhanced spectral events. Such a framework may be extendable beyond DNA to other heterogeneous biomolecular systems, including mRNA, peptides, microproteins, post-translational modifications, and conformationally dynamic biomolecules, where chemically meaningful states are often sparse, short-lived, and hidden within complex single-molecule SERS trajectories(*52–54*).

## METERIALS AND METHODS

### Materials

Citrate-stabilized, nonfunctionalized gold nanoparticles (AuNPs; 50 nm diameter, 3.5 × $10^{10}$ particles $ml^{-1}$) were purchased from Sigma-Aldrich. The AuNPs were supplied in 0.1 mM PBS and stabilized with citrate. Dry DNA oligonucleotides, including 5′-AAAAAAA-3′, 5′-ACCCCCA -3′, 5′-ATTTTTA -3′, 5′-AGGGGGA-3′, 5′-ACACACA-3′, 5′-AGAGAGA-3′, 5′-ATATATA-3′, 5′-TCTCTCT-3′, 5′-TGTGTGT-3′,5′-ACGCGCA-3′,5′-ATCGACTG-3′ and 5′-TACAAGTAAAG-3′, were purchased from Sigma-Aldrich and used without further purification unless otherwise stated. Silicon wafers coated with 100-nm $Si_3N_4$ membranes on p-type Si substrates were purchased from MicroChemicals GmbH. All aqueous solutions were prepared using ultrapure water.

### Instrumentation

SERS spectra were acquired using a Thermo Fisher DXR2xi Raman Imaging Microscope and collected with Andor Solis software. Nanowell arrays were fabricated by focused ion beam milling using an FEI Helios FIB system. Scanning electron microscopy images of the nanowells and assembled particle-in-well structures were acquired using a Sigma HD VP field-emission SEM. Gold films were deposited using a Q150T ES sputter coater. Capillary-assisted particle assembly was performed using a custom setup consisting of a power supply, a three-dimensional printed holder, a PID temperature controller, and a Peltier heater. Dynamic light scattering measurements were performed using a Zetasizer Nano. Spectral processing and data analysis were carried out in MATLAB R2023a and VScode.

### PIW-based SM-SERS measurements

DNA-loaded AuNPs were entrapped in the nanowell arrays to form particle-in-well plasmonic hotspots. Time-resolved SM-SERS measurements were performed on selected nanoarrays using a 785-nm excitation laser. Spectra were collected with a 50× objective, a laser power of 18–22 mW, an exposure time of 0.1 s per spectrum, a full-range resolution grid, and a 50-μm slit. For each time-resolved trajectory, 500 consecutive spectra were acquired using Andor Solis. These time series were used to monitor dynamic single-molecule spectral fluctuations arising from local DNA–hotspot interactions.

### Spectral preprocessing

Raw SERS spectra were preprocessed before downstream peak assignment and statistical analysis. First, spectra were clipped to the Raman-shift region of interest. For SM-SERS trajectory analysis analysis, the 300–1400 $cm^{-1}$ region was used. Cosmic-ray artifacts were removed using a custom MATLAB function, remove_cosmic_rays, which detects and suppresses sharp spike-like features exceeding a defined intensity threshold.

### Peak detection and assignment

After preprocessing, Raman peaks were detected using the MATLAB findpeaks function. Peak assignment was performed using a reference database constructed from literature-reported vibrational modes of DNA bases, DNA backbone features, and citrate-related signals. Peaks were assigned by considering their Raman shift positions together with peak width, height, prominence, and allowed shift tolerance. Because single-molecule SERS spectra are strongly affected by adsorption geometry, molecular orientation, local electromagnetic field distribution, and transient hotspot coupling, peak assignments were treated as tentative rather than absolute. Recurrent peak regions were therefore interpreted as local molecular-state signatures instead of direct one-to-one matches to isolated-base reference spectra.

### Data analysis

Processed spectra were analyzed in Visual Studio Code using Python 3.8 and PyTorch. Time-resolved trajectories were used for molecular-state classification, dwell-length analysis, entropy calculation, switch-rate analysis, transition-matrix construction, and resampling-based robustness evaluation. All downstream analyses were performed on probability-filtered molecular-state assignments to ensure that trajectory-level statistics were primarily based on high-confidence SM-SERS spectral states.

## Supplementary Materials

**The PDF file includes:**

Supplementary Text

Figs. S1 to S46

Tables S1 to S27

References

**Acknowledgments:** We thank Z. Wu for suggestions regarding data-preprocessing. We thank Z. Chen and M. Yaltaye for discussions. **Fundings:** This work was supported by Marie Skłodowska-Curie Actions (grant no. 101126602); the National Natural Science Foundation of China (grant nos. W2521040, 82272054 and 22595410); the Science and Technology Commission of Shanghai Municipality (grant nos. 24DIPA00300, 24490710800, and 24490790900); and the Research Council of Finland mobility funding (grant no. 366158).

**Author contributions:** Conceptualization: K.Z., J.Y., Z.C., and J.H. Methodology: K.Z., J.Y., Z.C., and J.H., Investigation: K.Z, P.X., Y.Z., and J.H. Formal analysis: K.Z., P.X., H.G., and E.A. Validation: K.Z., P.X., Y.Z. and E.A. Data curation: K.Z., H.G., and E.A. Software: K.Z., H.G., and E.A. Resources: J.Y., Z.C., and J.H., Visualization: K.Z., Y.Z., and J.H. Supervision: J.H. Project administration: K.Z., Z.C., and J.H. Funding acquisition: K.Z., J.Y., Z.C., S.L., L.C., and J.H. Writing—original draft: K.Z. Writing—review and editing: Z.C., J.Y., S.L., L.C. and J.H.

**Competing interests:** the authors declare that they have no competing interests. **Data, code, and materials availability:** All data and code needed to evaluate and reproduce the results in the paper are present in the paper and/or the Supplementary Materials.